\documentclass{appolb}
\usepackage{amsmath,amssymb,epsfig,graphicx}

\begin{document}
\title{Jet reshaping in heavy-ion collisions%
\thanks{Presented at the 37th International Symposium on Multiparticle Dynamics, LBNL, August 2007.}%
}
\author{Carlos A. Salgado\footnote{carlos.salgado@cern.ch}
\address{Dipartimento di Fisica, Universit\`a di Roma "La Sapienza" \\
and INFN, Roma, Italy\\ and\\
Departamento de F\'\i sica de Part\'\i culas and IGFAE, \\ Universidade de Santiago de Compostela, Spain}
}
\maketitle
\begin{abstract}
We propose a new implementation of medium effects in jet structures in which a modification of the splitting function is included at every step in the typical final state parton shower. Although the main application of this new formalism will be at the LHC, it is interesting that, in the presence of a trigger bias to small number of splittings, non-trivial angular dependences could appear with shapes similar to those measured experimentally at RHIC in high-$p_T$ particle correlations.
\end{abstract}

The presence of a medium is known to modify the branching evolution of the quarks or gluons produced at high-$p_T$ by a hard process. The corresponding situation in the vacuum is well known and allows a probabilistic interpretation in which every branching -- dictated by a splitting function, $P(z)$ -- reduces the initial virtuality of the produced parton until some given scale ${\cal O}$(1GeV) is reached and the process is stopped. In this way, the DGLAP evolution of the fragmentation function can be written at LO as
\begin{equation}
D(x,t)=\Delta(t)D(x,t_0)+\Delta(t)\int_{t_0}^t \frac{dt_1}{t_1}
\frac{1}{\Delta(t_1)} \int \frac{dz}{z} \, P(z)
D\left(\frac{x}{z},t_1\right).
\label{eq:dglapsud}
\end{equation}
The first term on the right-hand side in this expression corresponds  to the contribution with no splittings between $t_0$ and $t$ while the second one gives the evolution when some finite amount of radiation is present. The evolution is controlled by the Sudakov form factor
\begin{equation}
\Delta(t)=\exp{\left[-\int_{t_0}^{t} {dt^\prime \over t^\prime}
\int dz {\alpha_s(t^\prime,z)
\over 2 \pi} P(z,t^\prime)\right]},
\label{eq:sudakovs}
\end{equation}
with the interpretation of the probability of no resolvable branching between the two scales $t$ and $t_0$. 

The definition of the Sudakov form factors and its probabilistic
interpretation depend on the cancellation of the different divergencies
appearing in the corresponding Feynman diagrams. Although such cancellation has never been proved on general grounds for partons re-scattering in a medium,  it has been found in \cite{Wang:2001if} that, under certain assumptions, all the medium effects can be included in a redefinition of the splitting function
\begin{equation}
P^{\rm tot}(z)= P^{\rm vac}(z)+\Delta P(z,t),
\label{eq:medsplit}
\end{equation}
where we have labeled as "vac" the corresponding vacuum splitting function. The main assumption to arrive at (\ref{eq:medsplit}) is the independence of the multiple gluon emission when re-scattering with the nuclei is present so that an exponentiation of the splittings is possible. This possibility was exploited in \cite{Armesto:2007dt} where the additional term in the splitting probability is just taken from the medium-induced gluon radiation by comparing the leading contribution in the vacuum case.
\begin{equation}
\Delta P(z,t)\simeq \frac{2 \pi  t}{\alpha_s}\, 
\frac{dI^{\rm med}}{dzdt} ,
\label{medsplit}
\end{equation}
Implementing (\ref{eq:medsplit}) and (\ref{medsplit}) into (\ref{eq:dglapsud}) the medium-modified fragmentation functions can be computed -- see also \cite{Borghini:2005em} for a related approach. Only the initial conditions of the evolution need to be specified. In \cite{Armesto:2007dt} the KKP set of FF \cite{Kniehl:2000fe} was used for the vacuum as well as for the medium at the initial scale $Q^2_0$, i.e. $D^{\rm med}(x,Q^2_0)=D^{\rm vac}(x,Q^2_0)$. In this model all the medium-effects are built during the evolution. The motivation for this ansatz is the following: in hadronic collisions, particles produced at high enough transverse momentum hadronize outside the medium. So, this assumes that the non-perturbative hadronization is not modified by the medium, whose effect is only to modify the perturbative associated radiation. All present radiative energy loss formalisms rely on this assumption.

\begin{figure}
\begin{minipage}{0.5\textwidth}
\begin{center}
\includegraphics[width=0.8\textwidth]{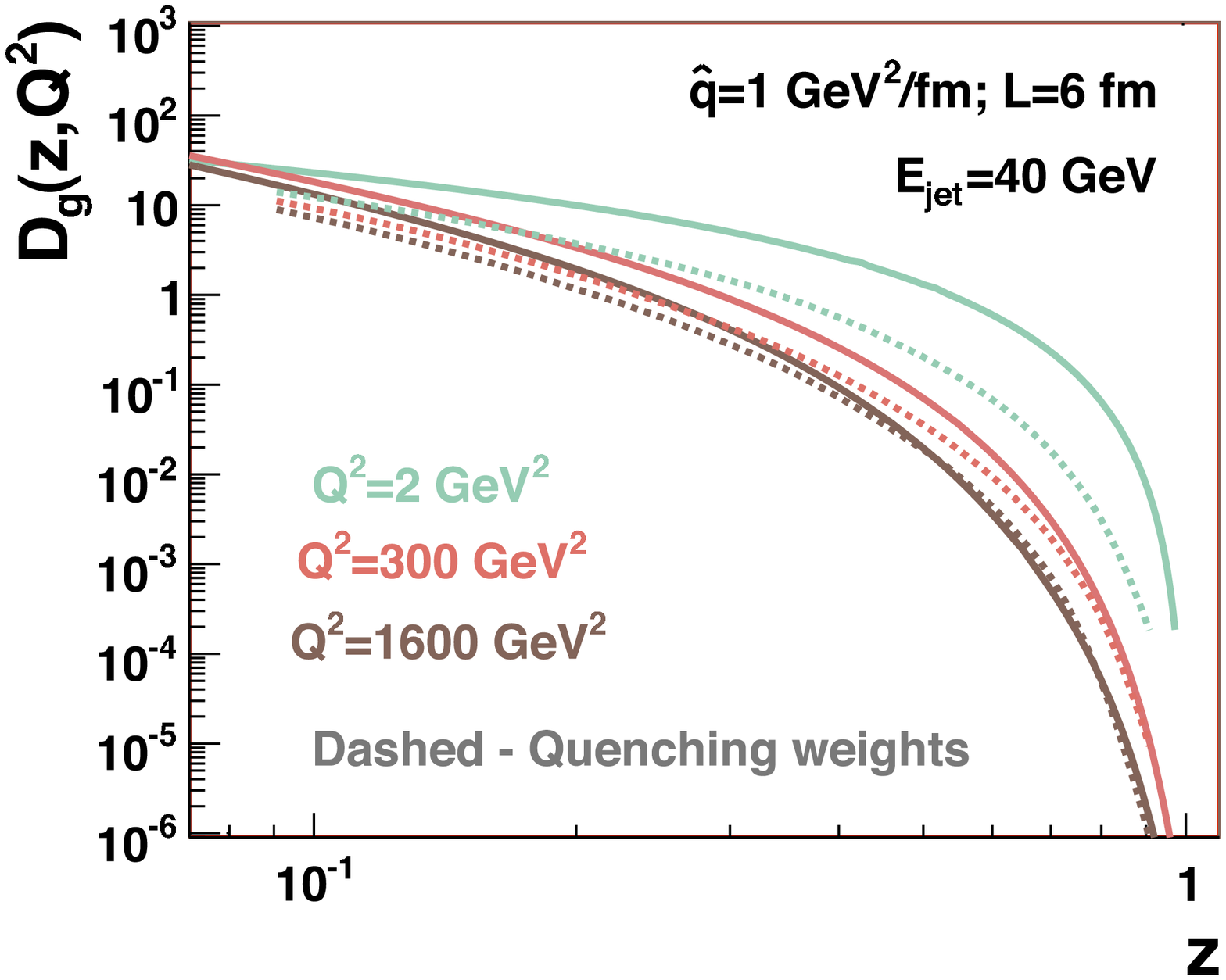}
\end{center}
\end{minipage}
\hfill
\begin{minipage}{0.5\textwidth}
\begin{center}
\includegraphics[width=0.85\textwidth]{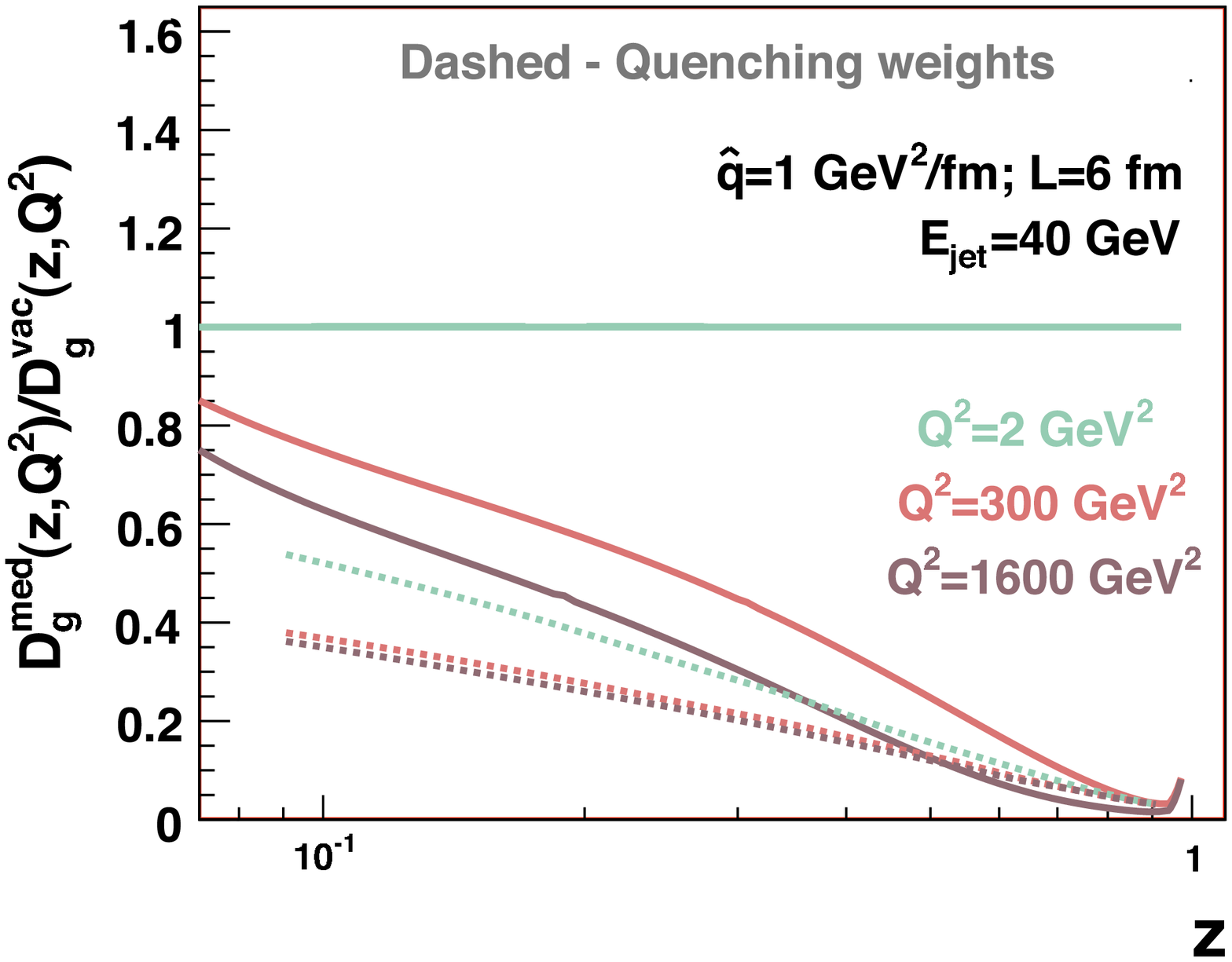}
\end{center}
\end{minipage}
\caption{Left: Fragmentation function for gluons onto pions computed with the medium-modified evolution described in the text (solid lines) and through the standard convolution with the quenching weights (dashed lines). Right: Ratio of the fragmentation function for gluons onto pions in a medium over that in the vacuum.}
\label{fig:ff}
\end{figure}

Interestingly, the fact that the medium-induced gluon radiation, $I^{\rm med}$, is infrared and collinear finite allows for a simplification of this formulation, valid when $E\sim Q\gg 1$. Under these conditions, eqs. (\ref{eq:dglapsud}) and (\ref{eq:sudakovs}) reduce to the well known quenching weights \cite{Salgado:2003gb} previously used in most jet quenching phenomenology. In Fig. \ref{fig:ff} we plot the fragmentation functions in the presence of a medium for different values of $\hat q$ and $Q^2$ by including the modification at every individual splitting, eqs. (\ref{eq:dglapsud})--(\ref{medsplit})  \cite{Armesto:2007dt}. Also plotted in the same figure is the comparison with the corresponding FF using the standard quenching weights \cite{Salgado:2003gb}. Clearly, both procedures will provide similar descriptions of the experimental data on one-particle inclusive suppression  \cite{Armesto:2007dt} but the new approach allows for a clean implementation of the effects for less inclusive measurements as jet structures or particle correlations.

The most general case of a complete parton shower implementation with nuclear effects has not been computed yet. There is a simplification, in a very specific situation, in which the radiated gluons have energy $\omega\lesssim 2\hat q^{1/3}$ where the medium-induced gluon radiation can be approximated by \cite{Polosa:2006hb}
\begin{equation}
\frac{dI^{\rm med}}{d\omega dk_\perp^2}\simeq\frac{\alpha_s C_R}{16\pi}\, L\, \frac{1}{\omega^2}.
\label{eq:migrap}
\end{equation}
Keeping only the first term in the expansion of (\ref{eq:dglapsud}) we obtain the probability of just one splitting, which, inserting (\ref{eq:migrap}) and making the change to laboratory azimuthal angle for a jet produced at $\eta=0$ reads
\begin{equation}
\frac{d{\cal P}(\Phi,z)}{dz\,d\Phi}\Bigg\vert_{\eta=0}=
\frac{\alpha_s C_R}{16\pi^2}\,E\,L\,\cos\Phi
\exp\left\{-E\,L\,\frac{\alpha_s C_R}{16\pi}\cos^2\Phi\right\}
\label{eq:splitlab}
\end{equation}

\begin{figure}
\begin{minipage}{0.49\textwidth}
\begin{center}
\includegraphics[width=0.75\textwidth,angle=-90]{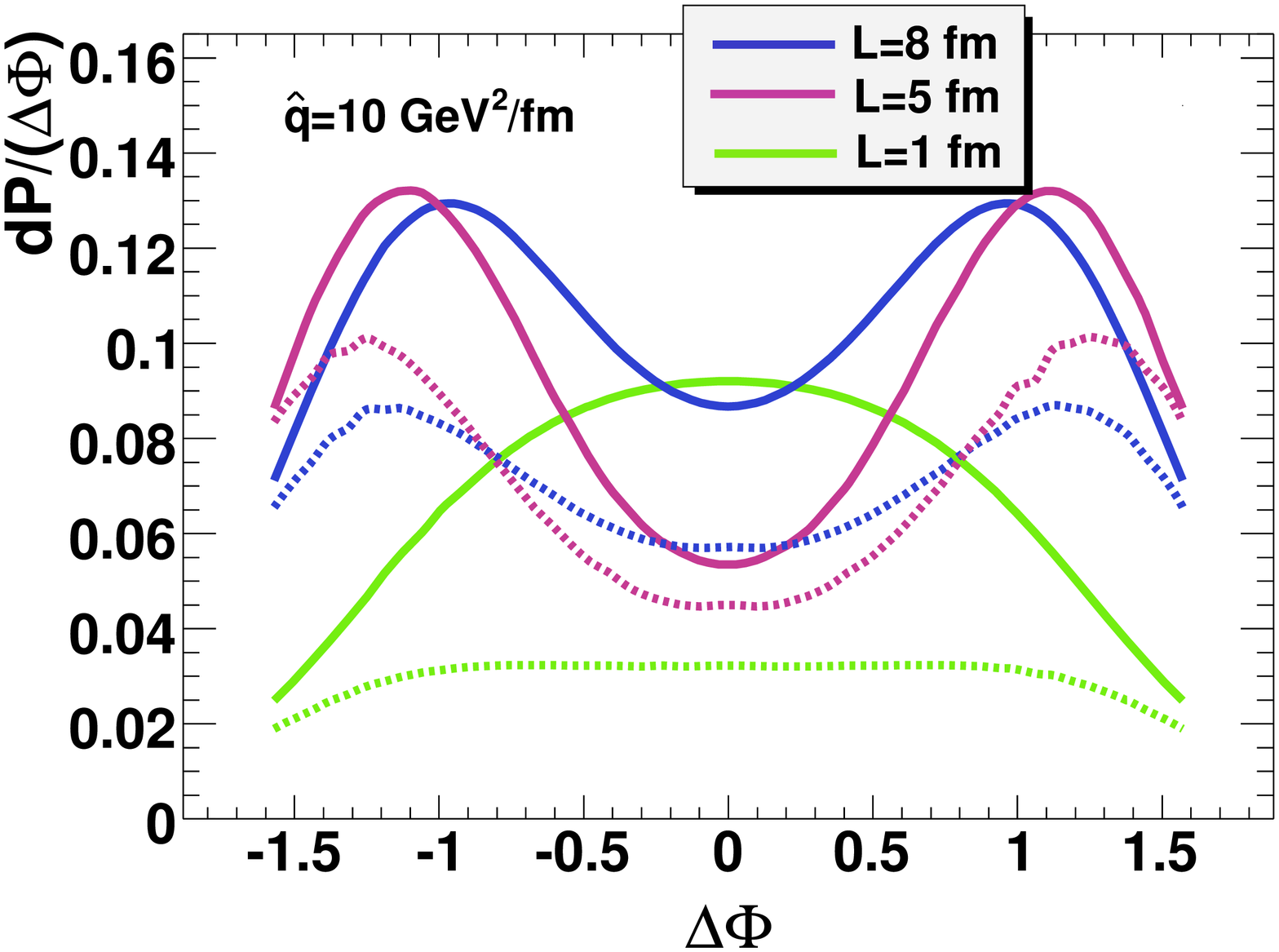}
\end{center}
\end{minipage}
\hfill
\begin{minipage}{0.49\textwidth}
\begin{center}
\includegraphics[width=0.75\textwidth,angle=-90]{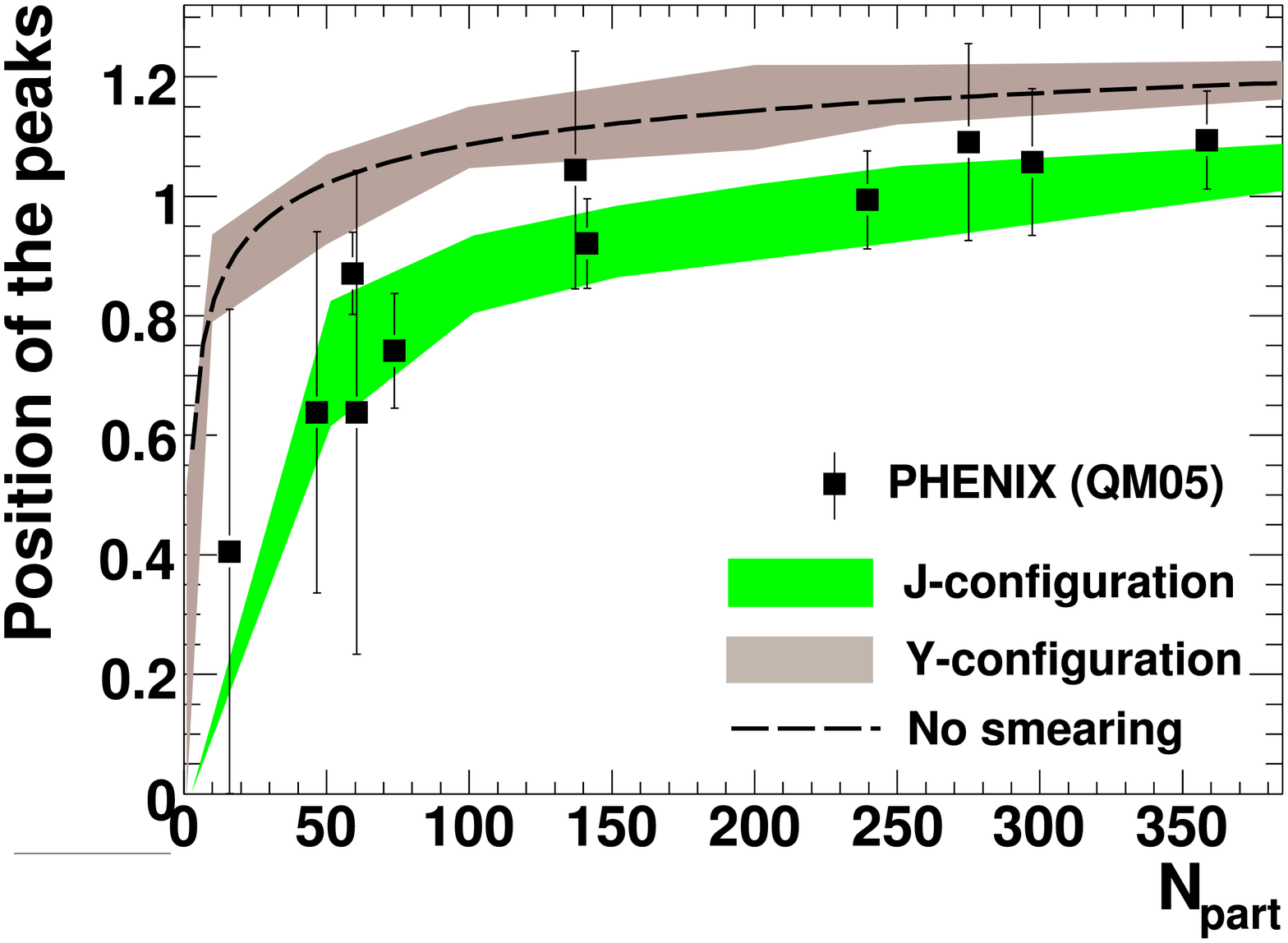}
\end{center}
\end{minipage}
\caption{(Left) The probability of just one splitting (\protect\ref{eq:splitlab}) as a function of the laboratory azimuthal angle $\Delta\Phi$ for a gluon jet of $E=7$ GeV. A smearing in $\eta$ and $\Phi$ is included. (Right) Position of the peaks with PHENIX data \cite{Adler:2005ee}. Figures \protect\cite{Polosa:2006hb}.
\label{fig:ndist}}
\end{figure}

Eq. (\ref{eq:splitlab}) presents a non-trivial angular structure in which, due to the Landau-Pomeranchuk-Migdal suppression of radiation at small angles, a dip is observed in the direction of the original jet in contrast with the known situation in the vacuum, where collinear singularities make the jet shapes to be well collimated around $\Phi=0$. Similar structures has been observed experimentally by two particle correlations at moderate values of the transverse momentum \cite{Adler:2005ee}. Although the calculations presented here lack of a realistic implementation, including multiple splittings and hadronization, it is encouraging to see a perturbative mechanism producing such a non-trivial shape \cite{Polosa:2006hb}. In order to make a rough comparison with experimental data, we just smeared the distribution in $\eta$ and $\phi$ to mimic the experimental situation. The corresponding distributions are plotted in Fig. \ref{fig:ndist}.

\section*{Acknowledgements}
CAS is supported by the FP6  of the European Community under the contract MEIF-CT-2005-024624.

\end{document}